\documentclass[11pt,a4paper]{article}       

\usepackage[numbers]{natbib}
\pdfoutput=1
\usepackage{ dsfont }
\usepackage{ stmaryrd }
\usepackage{ amssymb }
\usepackage{graphicx}
\usepackage{amsmath}
\usepackage{subcaption}
\usepackage{multirow}
\usepackage{multicol}
\usepackage{tabularx}
\usepackage{diagbox}
\usepackage{ mathrsfs }

\usepackage{pgfplots}
\usepackage{enumitem}
\usepackage{titlesec}
\usepackage{ulem} 
\usepackage{mdframed}
\usepackage{color}

\usepackage{tikz}
\usetikzlibrary{snakes}
\usetikzlibrary{patterns}

\pgfplotsset{
        compat=newest, 
        every axis/.append style={
        scaled y ticks = false, 
        scaled x ticks = false, 
        y tick label style={/pgf/number format/.cd, fixed, fixed zerofill,
                            int detect,1000 sep={\;},precision=3},
        x tick label style={/pgf/number format/.cd, fixed, fixed zerofill,
                            int detect, 1000 sep={},precision=3}
    }
}

\usepackage[pdftex,colorlinks=true,urlcolor=blue,pdfstartview=FitH]{hyperref}
\DeclareGraphicsExtensions{.pdf,.pdftex_t,.png,.jpg} 
\graphicspath{{../images/}{.}{../../images/}{./images/}}

\titleclass{\subsubsubsection}{straight}[\subsection]
\newcounter{subsubsubsection}[subsubsection]
\renewcommand\thesubsubsubsection{\thesubsubsection.\arabic{subsubsubsection}}

\titleformat{\subsubsubsection}
  {\normalfont\normalsize\bfseries}{\thesubsubsubsection}{1em}{}
\titlespacing*{\subsubsubsection}
{0pt}{3.25ex plus 1ex minus .2ex}{1.5ex plus .2ex}

\makeatletter
\def\toclevel@subsubsubsection{4}
\def\l@subsubsubsection{\@dottedtocline{4}{7em}{4em}}
\makeatother

\newmdtheoremenv{definition}{Definition}
\newtheorem{lemma}{{\bf Lemma~}}
\newtheorem{theorem}{Theorem}
\newtheorem{remark}{Remark}
\newtheorem{conjecture}{Conjecture}
\newtheorem{corollary}{Corollary}
\newcommand{\qed}{$\square$}

\newcommand{\vn}[1]{$\cal{N}$$^V${$(#1)$}}
\newcommand{\en}[1]{$\cal{N}$$^E${$(#1)$}}
\newcommand{\gn}[1]{$\cal{N}$$^G${$(#1)$}}
\newcommand{\nervgraph}{$\cal{N}$$^G$}
\newcommand{\nervvertex}{$\cal{N}$$^V$}

\begin{document}

\title{Dynamic Graphs Generators Analysis: an Illustrative Case Study}
\author{Vincent Bridonneau \and Frédéric Guinand \and Yoann Pigné}

\maketitle

\begin{abstract}
In this work, we investigate the analysis of generators for dynamic graphs, which are defined as graphs whose topology changes over time. We introduce a novel concept, called "sustainability," to qualify the long-term evolution of dynamic graphs. A dynamic graph is considered sustainable if its evolution does not result in a static, empty, or periodic graph. To measure the dynamics of the sets of vertices and edges, we propose a metric, named "Nervousness," which is derived from the Jaccard distance.
As an illustration of how the analysis can be conducted, we design a parametrized generator, named D3G3 (Degree-Driven Dynamic Geometric Graphs Generator), which generates dynamic graph instances from an initial geometric graph. The evolution of these instances is driven by two rules that operate on the vertices based on their degree. By varying the parameters of the generator, different properties of the dynamic graphs can be produced.
Our results show that in order to ascertain the sustainability of the generated dynamic graphs, it is necessary to study both the evolution of the order and the Nervousness for a given set of parameters.
\end{abstract}

\newpage 

\tableofcontents

\newpage

\section{Introduction}
\label{sec:intro}

Nature and human societies offer many examples of systems composed of entities that interact, communicate or are just connected with each other. 
The Internet, a transportation network, a swarm of robots, an ant colony, a social network, a urban network, or a crowd are some examples \cite{boccaletti.et.al.2006}.  

Graphs are certainly one of the best formalisms for modeling them.
Every vertex in the graph models one entity. 
A link is added between two vertices when a particular condition about the corresponding entities is verified. 
For instance: two persons are talking to each other, a predator catches a prey, two playing cards are in the same hand, a virus passes from one individual to another, two actors perform in the same play, etc.
The semantic of the interaction, communication or connection is proper to the system.

During last two decades, many works have been dedicated to the study of networks modeling these systems.
It has been shown that, unlike classical, regular or random graphs, graphs modeling complex real systems present specific statistical properties, leading researchers to introduce the term of complex networks for naming them. 
Among main characteristics that were highlighted are the small-world and the scale-free properties. 
The small-world property was discovered many years ago, in the 60s, when Stanley Milgram imagined and conducted the "small-world problem" in order to measure, through postal mail, the number of intermediaries between any two persons in US \cite{milgram.1967,travers.milgram.1977}.  
The scale-free property was also observed quite a long time ago for some datasets.

One fundamental question motivates researchers working on these networks: "which mechanisms might enable such properties to appear in these networks?"
But these networks did not appear as they are today.
Thus, finding mechanisms means designing a generative or building process able to produce graphs owning these properties. 
In 1998, Watts and Strogatz proposed a way of generating graphs owning the small-world property. 
Starting from a regular lattice the process randomly rewires part of the network's connections \cite{watts.strogatz.1998}.
One year later, in 1999, Barab{\`a}si and Albert designed an algorithm for generating graphs with both the small-world and the scale-free properties  \cite{barabasi.albert.1999}.
This process generates growing networks. 
At each time step a new vertex is added to the graph and is more likely linked to high degree vertices.  
This mechanism or rule is known as "preferential attachment". 
In both cases, the approach used consists in designing processes based on appropriate mechanisms to generate graphs with the desired properties.
The approach may be qualified as "on purpose design". \\

The work developed in this report addresses the opposite problem. 
Given a generative process, questions asked are: 
\begin{itemize} 
\item how the dynamics of generated graphs can be characterized?
\item what metrics might be used for that purpose, and how to compute them?
\item from the point of view of dynamics, is it possible to classified or gathered generators into classes or families? 
\end{itemize}

In this ongoing work, not all questions are addressed. 
But we hope it will be a milestone for carrying out analysis of dynamic graphs generators. 
For this, in Section \ref{sec:model} a generic diagram of generators is presented and discussed. 
It is followed by the definition of a novel notion based on a specific metric, both targeting the dynamics of the graphs. 
Finally, the Degree-Driven Dynamic Geometric Graph Generator (D3G3) is presented. 
D3G3 is a parameterized generator and according  to the parameters, it can produce a wide variety of dynamics. 
It will be used as a case study.  
A first global analysis of the generated graph families is performed in Section \ref{sec:theoreticalanalysis}. 
Section \ref{sec:segments} focuses on specific values of the parameters and present a rigorous analysis of the evolution of the dynamics of the graph and of the likelihood of its sustainability.
A conclusion, drawing some perspectives and future investigations, closes temporarily this report.

\section{Definitions and Generative Model and Definitions}
\label{sec:model}

\subsection{Notations}

Consider two sets $A$ and $B$:
\begin{itemize}
\item $|A|$ refers to the number of elements of set $A$.
\item $A-B$ refers to the set of elements present in set $A$ and absent in set $B$. 
\item $\triangle$ operator: $A\triangle B$ is defined as $A \cup B - A \cap B$. 
For instance if $A=\{1,2,3,4,5\}$ and $B=\{4,5,6,7,8\}$ then $A \triangle B = \{1,2,3,6,7,8\}$
\item Consider $G$ a dynamic graph, $G_t = (V_t,E_t)$ denotes the state of the graph at time $t$, where $V_t$ is the set of vertices and $E_t$ is the set of edges. 
\item $|V_t|$ (resp. $|E_t|$) corresponds to the number of vertices (resp. edges) of graph $G_t$.
\item for simplifying notations in the document, $|V_t|$ is often denoted by $n_t$
\item $G=(V,E)$ is said to be a null graph if both $V = \emptyset$ and $E = \emptyset$. In the report such a null graph can also be called an empty graph. 
\end{itemize}

\subsection{Position of the work with respect to Temporal Networks}

Current definitions of temporal networks (TN) include \textit{time-varying graphs} \cite{santoro_time-varying_2011}, \textit{temporal networks} \cite{kempe_connectivity_nodate}, \textit{evolving graphs} \cite{ferreira_note_2002} etc.
They all define structures described by sequence of static graphs, ordered by a timestamp (e.g., $G=(G_i=(V_i, E_i))_{i \geqslant 0}$) where $i$ refers to the time step).
It is worth mentioning that TN definitions do not include information about the generative process.
Thus, the way the graph at time $t+1$ is obtained from the graph at time $t$ is not described.
In this report, the emphasis is precisely on the study of generative processes. 
This work is therefore positioned upstream of TN.
In the sequel graphs produced by generators are called dynamic graphs or simply graphs. 

\subsection{Generalities}

From a general point of view, a dynamic graph generator can be defined as a process with input data, that produces at each time step $t+1$ a new static graph $G_{t+1}$. It is produced from already generated static graphs $\{G_1,\ldots,G_t\}$ and possibly additional information.
Thus, the output of a dynamic graph generator is a flow of static graphs identified by time stamps. 
The time stamps may also corresponds to events, and in such a case, the time interval between two time stamps may be different. 
However, in this report, for sake of clarity, we consider integer time stamps. 
If the flow stops, for whatever reason (e.g. clock has been stopped, evolution process is finished) at step $T$, the set of generated static graphs $\{G_1,G_2,\ldots,G_T\}$ corresponds to a temporal network (TN). 

\begin{center}
\centering
\begin{tikzpicture}[-,thick,xscale=1.0,yscale=1.0]
\tikzstyle{process}=[rectangle,draw=black,fill=blue!10,text=black] 
\tikzstyle{input}=[rectangle,draw=black,fill=white,text=black]
\tikzstyle{output}=[rectangle,draw=black,fill=blue!50,text=white]

    \node[process, text width=3.2cm] (G) at (5,5) {Evolution Process $\{G_1,G_2,\ldots,G_t\}$};
    \node[input] (I) at (1,5) {Inputs};
    \node[process] (CK) at (4,7) {Clock};
    \node[output] (TN) at (9,5) {$G_{t+1}$};
    \node (TIME) at (3.5,6.2) {$t+1$};
    \node[process] (INFO) at (5,3.5) {Information};

    \draw[->,thick] (I.east) --  (G.west);
    \draw[->,thick] (CK) -- (4,5.6);
    \draw[->,thick] (G) -- (TN);
    \draw[->,thick] (INFO) -- (G);
    \draw[->,>=stealth,thick] (G.east) .. controls (10,6) and (6,8) .. (G.north)
        node[sloped, above,midway] {$G_{t+1}$};

\end{tikzpicture}
\end{center}

\subsection{Sustainability}
\label{sec:sustainability}

The goal of this section is to introduce a novel notion for qualifying the dynamics of a graph.
The only measurement of the order (or of the density) of the graph is not enough for qualifying its dynamics. 
For instance, if a dynamic graph becomes static, all vertices remain the same and the graph order does not change.
Conversely, if between two consecutive time steps all vertices are replaced by new ones, the order also remains the same, but the dynamics is different. 
Sustainability qualifies a dynamic graph that never becomes null or periodic (which includes static).
A graph owing the sustainability property is said sustainable. \\

\begin{definition}
\underline{Graph sustainability}\\
A dynamic graph $G$ is said sustainable if both Condition 1 and Condition 2 are not verified. 
$$
\begin{array}{ll}
   \mbox{Condition 1:} & \exists T \in \mathds{N}, \forall t \geqslant T, G_{t} = (\emptyset,\emptyset)\\
   \mbox{Condition 2:} & \exists T \in \mathds{N}~\mbox{and}~\exists k \in \mathds{N^*}, \forall t \geqslant T, G_t = G_{t+k}
\end{array}
$$
\end{definition}
\vspace{0.3cm}

Some well-known graph generators produce sustainable dynamic graphs. 
For instance generators of growing networks. 
Indeed, for all $t \in \mathds{N}$, $|V_t| > |V_{t-1}|$, and $G_t \neq (\emptyset,\emptyset)$.
For these generators, graph sustainability is obvious and does not require any analysis. 

Unlike these cases, some generators are based on mechanisms making the evolution of the vertices (and edges) more difficult to predict, and the dynamics is worth studying. 
For that purpose, we propose to consider a metric enabling a quantification of the dynamics.

\subsection{Nervousness}
\label{sec:nervousness}

This metric provides a way of measuring the dynamics of a graph.
Note that this metric is not new and is known as the Jaccard distance, which can be defined as one minus the coefficient of community as defined in \cite{jaccard_1912}.
However, in the context of dynamic graphs it seems to us more meaningful to call it nervousness. 
It is different from the burstiness which is defined at the node/edge level during the lifetime of the graph \cite{goh.barabasi.2008} and aims at representing the frequency of events occurring on each node/edge.
Nervousness metric aim is to capture the dynamics of creation and deletion of nodes and edges between two time steps at graph level. \\

\begin{definition}
\underline{Vertices Nervousness}:\\
Given a dynamic graph $G$, such that at time $t$ $G_t = (V_t,E_t)$.
We call {\bf vertices nervousness} at time $t$ and denoted by \vn{t}, the ratio:
\[
  \mbox{\vn{t}} = \frac{|V_{t+1} \triangle V_t|}{|V_{t+1} \cup V_t|} = \frac{|V_{t+1} \cup V_t - V_t \cap V_{t+1}|}{|V_{t+1} \cup V_t|}
\]
\end{definition}
\vspace{0.2cm}

This metric is complementary with the graph order measure.
Indeed, graph order can remain constant between two consecutive time steps although some vertices change. 
If all vertices are replaced, nervousness equals 1.
If all vertices are kept, nervousness is 0.
Similarly we define the edges nervousness as: \\

\begin{definition}
\underline{Edges Nervousness}:\\ Given a dynamic graph $G$, such that at time $t$ $G_t = (V_t,E_t)$. 
We call {\bf Edges Nervousness} at time $t$ and denoted by \en{t}, the ratio:
\[
  \mbox{\en{t}}= \frac{|E_{t+1} \triangle E_t|}{|E_{t+1} \cup E_t|}
\]
\end{definition}
\vspace{0.3cm}

At graph level, nervousness is defined as a couple:\\

\begin{definition}
\underline{Graph Nervousness}:\\ Given a dynamic graph $G$, the {\bf Graph Nervousness} at time $t$ is:
\[
 \mbox{\gn{t}}=(\mbox{\vn{t},\en{t}})
\]
\end{definition}
\vspace{0.2cm}

For illustrating these definitions, consider the following cases for a dynamic graph, from $t$ to $t+1$.
We denote $|V_t| = n_t$.  We also assume that between $t$ and $t+1$ the order remains the same, thus $|V_{t+1}| = n_{t+1} = n_t = n$.
\begin{itemize}
    \item if all vertices are replaced: \\
	$$\mbox{\vn{t}}= \frac{|V_t \triangle V_{t+1}|}{|V_t \cup V_{t+1}|} = \frac{2n}{2n} = 1$$
    \item if half of the vertices are replaced: \\
	$$\mbox{\vn{t}} = \frac{|V_t \triangle V_{t+1}|}{|V_t \cup V_{t+1}|} = \frac{3n/2-n/2}{3n/2} = \frac{n}{3n/2} = \frac{2}{3}$$
    \item if the vertices remain the same, the union of the sets is equal to their intersection thus:\\ 
	$$\mbox{\vn{t}}=0$$
\end{itemize}

When the order changes, for instance if all vertices are duplicated, thus $|V_{t+1}| = 2n_t = 2n$: \\
	$$\mbox{\vn{t}} = \frac{|V_t \triangle V_{t+1}|}{|V_t \cup V_{t+1}|} = \frac{2n-n}{2n} = \frac{1}{2}$$

Same results hold for edge nervousness. 
Dynamic graphs keeping the same set of vertices but with varying set of edges have a graph nervousness equal to: \gn{t} = (0,\en{t})\\

If \gn{t} = (\vn{t},0) with \vn{t} $ \neq 0$, then the set $V_{t+1}-V_t$ and the set $V_t - V_{t+1}$ contain only isolated nodes.

\subsection{Sustainability vs Nervousness}
\label{sec:sustainabilityvsnervousness}

Sustainability and nervousness are closely related. 
Sustainability describes a dynamic graph property while nervousness enables the measure of the evolution of vertices and edges sets between two consecutive time steps. 
When nervousness is null for both sets, the graph is static and thus does not have the sustainability property. 
Given a dynamic graph $G$, if for all $t \in \mathds{N}$, \vn{t} $\neq 0~\mbox{or}~\mbox{\en{t}} \neq 0$, then $G$ holds the sustainability property, except if $G$ is periodic. 

\subsection{D3G3: definition}

In this section we define a parameterized model generating families of dynamic graphs. 
An instance of the generative model is defined by a set of parameters. 
For studying the model, we analyze, according to the parameters set, the dynamic graphs families produced and rely on both the sustainability and the nervousness for that purpose. 

The generator has two types of inputs: a set of parameters, $S_p$, and an initial graph, called seed graph and denoted $G_0$. 
At each time step $t+1$ it produces, from the previous graph $G_t$, a new graph $G_{t+1}$ as illustrated on the figure. 

\begin{center}
\centering
\begin{tikzpicture}[-,thick,xscale=1.0,yscale=1.0]
\tikzstyle{process}=[rectangle,draw=black,fill=blue!10,text=black]
\tikzstyle{input}=[rectangle,draw=black,fill=white,text=black]
\tikzstyle{output}=[rectangle,draw=black,fill=blue!50,text=white]

    \node[process,text width=3cm,align=center] (G) at (6,5) {Evolution Process $G_t$};
    \node[input] (G0) at (1,4) {seed graph $G_0$};
    \node[input] (P) at (1,6) {$S_p=\{\mbox{parameters}\}$};
    \node[process] (CK) at (5,7) {Clock};
    \node (TIME) at (4.5,6.2) {$t+1$};
    \node[output] (TN) at (10,5) {$G_{t+1}$};
    
    \draw[->,thick] (P.east) -- (3,6) -- (3.5,5) -- (G.west);
    \draw[->,thick] (G0.east) -- (3,4) -- (3.5,5) -- (G.west);
    \draw[->,thick] (CK) -- (5,5.5);
    \draw[->,thick] (G) -- (TN);
    \draw[->,>=stealth,thick] (G.east) .. controls (10,6) and (7,8) .. (G.north)
        node[sloped, above,midway] {$G_{t+1}$};
\end{tikzpicture}
\end{center}

Graphs produced by D3G3 are geometric graphs. 
A geometric graph is defined by an euclidean space and a threshold $d$. 
For this study, without loss of generality we consider a 2D-unit-torus (i.e., a square $[0;1[^2$ where the two opposite sides are connected). 
Each vertex is characterized by a set of coordinates, such that given two vertices $u$ and $v$ it is possible to compute their euclidean distance: $dist(u,v)$. 
Given $V$ the set of vertices, the set of edges $E$ is defined in the following way:
$E=\{(u, v) \in V^2~|~\mbox{dist}(u, v) \leqslant d\}$\\

Graphs generated by D3G3 are produced thanks to an evolution process. 
This mechanism is parameterized by an initial graph (the seed graph) and by two transition rules driving the evolution of the graph between two consecutive time steps.
Apart from a random generator, no external decision or additional information is used by this mechanism. 
Rules are based on node degrees only and rely on a random generator for positioning new nodes in the 2D euclidean space.  
This leads to the name of the generator: \textit{Degree-Driven Dynamic Geometric Graphs Generator} or D3G3. \\

From now, graphs we are studying are referred to as sequences of static graphs $(G_t=(V_t,E_t))_{t\geqslant0}$, where $t\geqslant0$ is the time step. 
The initial graph, $G_0$ ($t=0$) is called the seed graph. \\

\begin{definition}
\underline{Degree Driven Dynamic Geometric Graph Generator}
\\
An instance of D3G3 is defined by an initial graph, a set of parameters and two rules:
\begin{itemize}[topsep=2pt]
\setlength\itemsep{-0.1em}
\item $G_0 \neq (\emptyset,\emptyset)$ the seed graph,
\item parameters:
    \begin{itemize}[topsep=0pt]
    \setlength\itemsep{-0.1em}
    \item $d \in ]0, \frac{\sqrt{2}}{2}[$ 
    \item $S_s$ a set of non-negative integers
    \item $S_c$ a set of non-negative integers
    \end{itemize}
\item rules applied on $G_t$ leading to $G_{t+1}$:
    \begin{itemize}[topsep=0pt]
    \setlength\itemsep{-0.1em}
    \item if $v \in V_t$, then $v \in V_{t+1}$ iff $\mbox{deg}(v) \in S_S$ (conservation rule)
    \item if $v \in V_t$ and if $\mbox{deg}(v) \in S_C$ then add a new vertex to $V_{t+1}$ with a random position in the unit-torus (creation rule)
    \end{itemize}
\end{itemize}
\end{definition}
\vspace{0.2cm}

Thus, evolution of the graph between two consecutive time steps $t$ and $t+1$, is driven by two rules applied to each vertex $v \in V_t$ simultaneously.
The first rule determines for a vertex $v \in V_t$ whether it is kept at step $t+1$ while the second rule concerns the possibility for a vertex $v \in V_t$ to create a new vertex in $V_{t+1}$ according to its degree. \\

\begin{definition}
\underline{Conserved/Create/Removed/Duplicated nodes}
\\
Let $t \geqslant 0$ and $G=(G_t)$ a D3G. Let $u \in V_t$ and $v \in V_{t+1}$, then
\begin{itemize}
    \item $u$ is said to be a conserved node iff $u \in V_t \cap V_{t+1}$.
    \item $u$ is said to be removed iff $u \in V_t - V_{t+1}$.
    \item $u$ is said to be a creator/creating node iff $\mbox{deg}(u) \in S_C$.
    \item $v$ is said to be a created node iff $v \in V_{t+1} - V_t$.
    \item $u$ is said to duplicate iff it is both a conserved and a creator node.
\end{itemize}
\end{definition}
\vspace{0.2cm}

Once created, a node never change its position.
Positions of created nodes do not depend on the creating nodes positions.
The position of a created node is chosen randomly and uniformly over the unit-torus.

\section{Theoretical Analysis} 
\label{sec:theoreticalanalysis}

While the model is very simple, it presents a wide variety of dynamics and long-term evolution.
According to $S_S$ and $S_C$ composition, several classes of dynamic behaviors have been identified.
These classes have been defined by computing two measures: the evolution of the order of the graph, and the evolution of the Graph Nervousness \nervgraph{}. 
Results are reported at the end of this section in Tables \ref{tab:cases} and \ref{tab:casessustainability}.

\subsection{Limit cases analyses}

Regarding the generative process, the obtained graphs depend on the threshold $d$, on the seed graph 
$G_0$ and on the two sets $S_C$ and $S_S$. Each combination of $d$, $S_C$ and $S_S$ does not necessarily lead to sustainable graphs.
$n_t$ denotes the order of $G_t$, and $n_0$ the order of the seed graph ($G_0$).

Some preliminary remarks and a first theorem help the analysis.

\begin{remark}\label{remark:deathnodenevercomeagain}
As soon as a node is removed from the graph, the same node will never appear again in the graph: \\
$\forall v \in V_{t-1}, v \notin V_t$, $\forall t' > t, v \notin V_{t'}$ 
\end{remark}

\begin{remark}\label{remark:noperiodic}
Graphs generated by D3G3 are not periodic (period $k > 1$).\\
$\forall G = (V,E), \nexists k>1,~\forall t,  V_{t+k} = V_t$
\end{remark}

Remark \ref{remark:noperiodic} is the direct consequence of Remark \ref{remark:deathnodenevercomeagain}. 
From these two remarks, Theorem \ref{theo:conclusionsustainable} can be formulated. 


\begin{theorem}\label{theo:conclusionsustainable}
Given $G=(V,E)$ a dynamic graph generated by D3G3, if its order and its nervousness are never equal to 0 then the graph is sustainable.\\
$\forall t>0$, if $n_t > 0$ and \vn{t} $>0$ the graph is sustainable.
\end{theorem}

\subsubsection{Case:  $S_S = \mathds{N}$ and $S_C = \mathds{N}$}

At each time step each vertex of the graph is both a conserved and a creator vertex, it is then duplicated. 
    Then, as soon as $G_0 \neq (\emptyset,\emptyset)$, the order of the graph increases exponentially: $n_t = 2^{t}n_0$ and thus
    $\lim_{t \rightarrow \infty} n_t = +\infty$.\\
    Consider an instance of dynamic graph produced by D3G3 with these parameters sets, is this graph sustainable? 
    The analysis of \nervvertex~leads to the conclusion.  
    Indeed, every vertex is duplicated, thus, $|V_{t+1}| = 2n_t$.
    From that, it comes $\mbox{\vn{t}} = \frac{|V_t \triangle V_{t+1}|}{|V_t \cup V_{t+1}|} = \frac{2n_t-n_t}{2n_t} = \frac{1}{2}$.
    In conclusion, for all $t$, (i) the graph is never null and (ii) \vn{t} $> 0$, thus from Theorem \ref{theo:conclusionsustainable} the graph is sustainable. 

\subsubsection{Case:  $S_S = \mathds{N}$ and $S_C = \emptyset$}

 If $S_S = \mathds{N}$ and $S_C = \emptyset$, then for all $t$, whatever the degree of any vertex, it is not a creator, hence no new nodes are added to the graph. 
In addition, all vertices have their degree in $S_S$, thus all nodes are conversed between $G_t$ and $G_{t+1}$.
So, $G_t = G_0$.
The nervousness \gn{t} $= (0,0)$, the graph is static: it is not sustainable.  

\subsubsection{Case: $S_S = \emptyset$  and $S_C = \mathds{N}$}

This case is the opposite situation of the previous one. 
All vertices are creators, but none is conserved. 
Hence, for all $t > 0$, if $v \in V_t$, $v \notin V_{t+1}$, but, as $deg(v) \in S_C$, the creation rule generates a new node that replaces $v$.
So, for all $t$, $n_t = n_0$. 
The graph is only dynamic, no subgraph is conserved between two consecutive time steps, all nodes are renewed.
From the nervousness point of view, this implies $V_t \cap V_{t+1} = \emptyset$ and $\mbox{\vn{t}} = \frac{|V_t \cup V_{t+1}|}{|V_t \cup V_{t+1}|}=1$, and the graph is sustainable. 

\subsubsection{Case: $S_S$ is a non-empty finite set and   $S_C = \emptyset$}

Note first that no vertex will be created in this situation since $S_C = \emptyset$. 
Thus for all $t$, $n_t \leq n_0$.
If $G_0$ is not null, then, several cases may occur:
\begin{enumerate}
\item first case: all vertices in $G_t$ have their degree in $S_S$: for all $v \in V_t$, $\mbox{deg(v)} \in S_S$. 
Then, all vertices are conserved, the order remains unchanged and the graph is static, $G_{t+1} = G_t$,  hence the graph is not sustainable, 
\item second case: opposite to the first one, for all $v \in V_t, \mbox{deg(v)} \notin S_S$, hence $G_{t+1}$ is null and the graph is not sustainable, 
\item last case: some vertices have their degree not belonging to $S_S$, they are removed and as a consequence $n_{t+1} < n_t$.
At each occurrence of this case, the order of the graph is strictly decreased by one, then this can happen at most $n_0$ times before the graph becomes empty unless it becomes static before. 
\end{enumerate}

As a consequence, for the limit case, the graph is not sustainable. 

\subsubsection{Case: $S_S = \emptyset$ and  $S_C$ is a non-empty finite set}

Note first that for all $t$, $V_{t+1} \cap V_t = \emptyset$, no vertex is conserved from $t$ to $t+1$. 
In addition, for all $t$, $n_{t+1} < n_t$, unless if for all $v \in V_t, \mbox{deg(v)} \in S_C$, then $n_{t+1} = n_t$.   
So the evolution of the graph order is non increasing. 
In addition, as $S_S = \emptyset$, \vn{t}$ = 1$, so the graph can never become static. 
The only way for the graph to be non sustainable is to become null. 
It is then not possible to conclude about the sustainability.
Indeed, consider the case for which at $t$, $0 < n_t \leq k+1$ and $S_C = [0,k]$ then for all $t' > t$, $n_{t'} = n_t$. 
Such a dynamic graph is sustainable. 
Conversely, if $S_C = \{k\}$, as soon as less than $k+1$ nodes have their degree equal to $k$ the graph becomes empty after two time steps. 

\subsubsection{Graph Order Increase}

For some specific parameters sets, the increase of the order of the graph asymptotically tends to zero, which has some effect on graph sustainability.  
This is the object of Theorem \ref{theo:low_increase}. 

\begin{theorem}\label{theo:low_increase}
Let $G=(G_t)_{t \geqslant 0}$, $S_S$ and $S_C$ such as $S_S$ or $S_C$ equals $\mathds{N}$ while the other one is finite and not empty, then the probability that graph order increases tends toward 0 as the graph order tends to infinity:
$P(n_{t+1} > n_t) \underset{\scriptscriptstyle {n_t \to +\infty}}{\to} 0$.
\end{theorem}

\noindent {\textbf{Proof}}
Let denote $P(n_{t+1} > n_t)$ the probability that graph order increases between step $t$ and step $t+1$ and $D=S_S \cap S_C$.
For a node $u \in V_t$, $u$ duplicates if and only if $\mbox{deg}(u) \in D$. 
\begin{align*}
P(n_{t+1} > n_t) &= P(\exists u, \mbox{deg}(u) \in D) \\
&= 1 - P(\forall u, \mbox{deg}(u) \notin D) \\
&= 1 - {(P(\neg D))}^{n_t }
\end{align*}
Where $P(\neg D)$ represents the probability a node does not satisfy the duplicating condition.
As positions of vertices are independent of each others:
\[
P(\neg D)=1-\sum_{k\in D}{n_t - 1 \choose k}p^k{(1-p)}^{n_t-k-1}
\]
Let $M = \max{D}$. For large values of $n_t$ and as $p \in ]0, 1[$:
\[\forall k \leqslant M, 0 < {n_t - 1 \choose k}p^k{(1-p)}^{n_t-1-k} \leqslant {n_t - 1 \choose M} {(1-p)}^{n_t-1-M}\]
As $0 < |D| \leqslant M+1$:
\begin{align} \label{eq:inequality_growth}
P(\neg D) \geqslant 1-\left({(M+1){n_t-1 \choose M}{(1-p)}^{n_t-1-M}}\right)^{n_t}
\end{align}
Knowing that
\[
\forall N, \forall k \leqslant N, {N \choose k} = \frac{N!}{k!(N-k)!} = \frac{1}{k!}{\displaystyle \prod_{i=N-k}^{N}(N-i)}
\]
and that $M$ is a constant not depending on $n_t$:
\[
{n_t - 1 \choose M}  \underset{\scriptscriptstyle {n_t \to +\infty}}{\sim} \frac{{n_t}^M}{M!}
\]
Rewriting the right side of \ref{eq:inequality_growth} leads to:
\begin{equation*}
A{n_t-1 \choose M}(1-p)^{n_t} \underset{\scriptscriptstyle {n_t \to +\infty}}{\sim} \frac{A}{M!}n_t^M(1-p)^{n_t}\textnormal{~~~where $A=\frac{M+1}{{(1-p)}^{M+1}}$} 
\end{equation*}
As $1-p \in ]0;1[$ is not depending on $n_t$,
$n_t^M(1-p)^{n_t}$ tends toward 0 as $n_t$ tends to infinity. Thus, $A{n_t - 1 \choose M}(1-p)^{n_t} \to 0$, thereby:
\begin{align*}
(P(\neg D))^{n_t} &\underset{\scriptscriptstyle {n_t \to +\infty}}{\sim}
\exp{(n_t\ln{(1-\frac{A}{M!}{n_t}^M(1-p)^{n_t}}))}\\
& \underset{\scriptscriptstyle {n_t \to +\infty}}{\sim}
\exp{(\frac{A}{M!}{n_t}^{M+1}(1-p)^{n_t})}
\end{align*}
From this, $(P(\neg D))^{n_t} \to 1$,
and hence we deduce the wanted theorem. \qed

\subsubsection{Case: $S_S = \mathds{N}$ and $S_C$ is a non-empty finite set}

As $S_S = \mathds{N}$, between two time steps, all nodes are conserved, thus, for all $t$, $V_t \subseteq V_{t+1}$.
The first consequence is that if $G_0$ is not a null graph, for all $t$, $G_t$ is not a null graph. 
Let us analyze the evolution of the order of the graph. 
Given $t$ such that, there exists a vertex $v$ such that $\mbox{deg(v)} \in S_C$, then $n_{t+1} > n_t$.
But, if for some $t'$, all $v \in V_{t'} , \mbox{deg(v)} \notin S_C$, then for all $t > t'$, $n_t = n_{t'}$ and the graph will be static. 
From Theorem \ref{theo:low_increase} 
$P(n_{t+1} > n_t) \underset{\scriptscriptstyle {n_t \to +\infty}}{\to} 0$ which implies $P(n_{t+1} - n_t = 0) \underset{\scriptscriptstyle {n_t \to +\infty}}{\to} 1$.
As all vertices are conserved between two time steps:
\[
P(|V_{t+1}-V_t| = 0) \underset{\scriptscriptstyle {n_t \to +\infty}}{\to} 1
\]
This leads to this limit on the probability the graph becomes static:
\[
 P(G_t~\mbox{is static}) \underset{\scriptscriptstyle {n_t \to +\infty}}{\to} 1\\
\]

Hence, for this limit case, we can conclude that the graph is asymptotically non sustainable.  

\subsubsection{Case: $S_S$ is a non-empty finite set and   $S_C = \mathds{N}$}

For all $t$, for all $v \in V_t$, either $v$ is duplicated (if $\mbox{deg(v)} \in S_S$) or $v$ is not conserved but is a creator.
In both cases, between two consecutive time steps each vertex generates the creation of a new one. 
Thus for all $t$,  \vn{t}~$>~0$ and $n_t~>~0$, hence, from remark 3, the graph is sustainable. 

\subsubsection{Summary of results}

\begin{table}[!h]
    \centering
    \begin{tabular}{|c|c|c|c|} \hline
    \diagbox{$S_S$}{$S_C$} & $\mathds{N}$ & finite set & $\emptyset$ \\ \hline
     \multirow{3}{*}{$\mathds{N}$} 
                & $\forall t, n_t = 2^tn_0$        & $\forall t, n_{t+1} \geqslant n_t$         & $\forall t, G_t = G_0$ \\ 
                & $\forall t, \mbox{\vn{t}} = 0.5$ & $\forall t, 0 \leq \mbox{\vn{t}} \leq 0.5$ & $\forall t, \mbox{\gn{t}} = (0, 0)$ \\ 
                &                                  & $\lim_{n_t \rightarrow \infty} P(n_{t+1} > n_t)$ = 0 & \\ \hline

    \multirow{2}{*}{finite} 
        & $\forall t, n_{t+1} \geqslant n_t$ & General  
        & $\forall t, n_{t+1} \leqslant n_t$ \\
       & $\forall t, 0.5 \leq \mbox{\vn{t}} \leq 1$ & cases & $\lim_{t \rightarrow \infty} n = \mbox{constant}$\\ 
       set & $\lim_{n_t \rightarrow \infty} P(n_{t+1} > n_t) = 0$ & (see Section \ref{sec:generalcases})  & $\lim_{t \rightarrow \infty} \mbox{\gn{t}} = (0, 0)$ \\ \hline
        
    \multirow{3}{*}{$\emptyset$} & $\forall t, n_{t+1} = n_t$  & $\forall t, n_{t+1} \leqslant n_t$ & \multirow{3}{*}{$\forall G_0, G_1=(\emptyset, \emptyset)$} \\ 
    & $\forall t, \mbox{\gn{t}} = (1, 1)$ &  $\forall t, V_t \neq \emptyset \implies \mbox{\gn{t}} = (1,1)$ & \\  \hline
    \end{tabular}
    \caption{Order and Nervousness evolution for the different cases. $n_t$ denotes the order of graph $G_t$, \vn{t} its vertices nervousness and \gn{t} the graph nervousness.}
    \label{tab:cases}
\end{table}
Stemmed from these results, sustainability property of dynamic graphs can be established.
The results are reported on Table \ref{tab:casessustainability}.

\begin{center}
    \begin{table}[!h]
    \centering
    \begin{tabular}{|c|c|c|c|} \hline
    \diagbox{$S_S$}{$S_C$} & $\mathds{N}$ & finite set & $\emptyset$ \\ \hline
     \multirow{2}{*}{$\mathds{N}$} & Sustainable & Asymptotically  & Non sustainable \\
	                                            &                     & non sustainable & \\ \hline 
    \multirow{1}{*}{finite}       & Sustainable & General cases  & Non sustainable \\
                             set          &                    & (see Section \ref{sec:generalcases}) & \\ \hline
    \multirow{2}{*}{$\emptyset$} & Sustainable & Depends on  & Non sustainable\\ 
    &  & the parameters  & \\ \hline
    \end{tabular}
    \caption{Sustainability of dynamic graphs according to parameters sets of D3G3.}
    \label{tab:casessustainability}
\end{table}
\end{center}

\subsection{General Cases} 
\label{sec:generalcases}

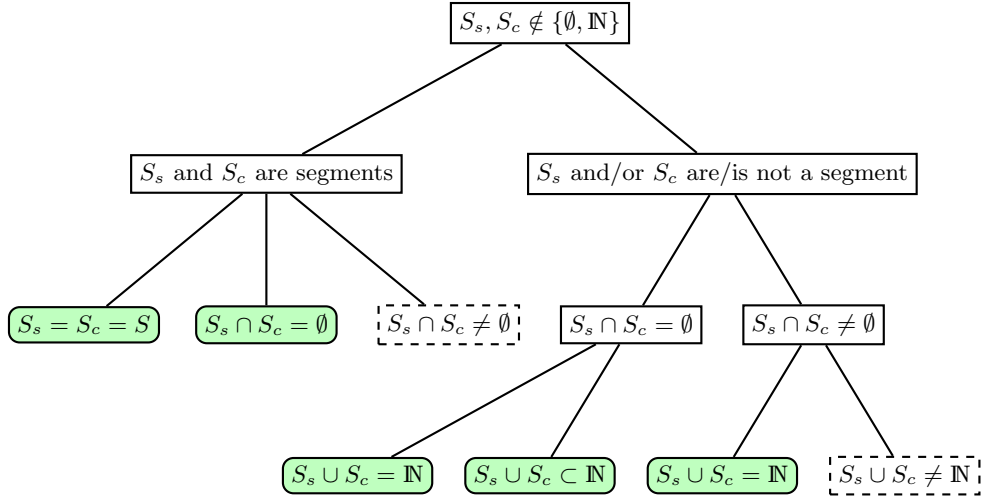
\begin{figure}[!h]
    \centering
    \begin{tikzpicture}[-,thick,xscale=0.4,yscale=0.4]
  \tikzstyle{sommet}=[rectangle,draw=black,fill=white,text=black]
  \tikzstyle{open}=[dashed,draw=black,text=black]
  \tikzstyle{ok}=[rectangle,rounded corners,draw=black,fill=green!25,text=black]
  \tikzstyle{arete}=[-]

  \begin{footnotesize}
    
    \node[sommet] (CG) at (12,13) {$S_s,S_c \notin \{\emptyset,\mathds{N}\}$};
    \node[sommet]  (SEG)   at (3,8) {$S_s$ and $S_c$ are segments};
    \node[sommet]  (NONSEG)   at (18,8) {$S_s$ and/or $S_c$ are/is not a segment};
    \draw (CG) -- (SEG);
    \draw (CG) -- (NONSEG);

    
    \node[ok]  (SEGEGAUX) at (-3,3) {$S_s = S_c = S$}; %
    \node[open] (SEGINTERNONVIDE) at (9,3) {$S_s \cap S_c \neq \emptyset$};
    \node[ok] (SEGINTERVIDE) at (3,3) {$S_s \cap S_c = \emptyset$};
    
    \draw (SEG) -- (SEGEGAUX);
    \draw (SEG) -- (SEGINTERNONVIDE);
    \draw (SEG) -- (SEGINTERVIDE);

     \node[sommet] (ENSINTERVIDE) at (15,3) {$S_s \cap S_c = \emptyset$};
     \node[sommet] (ENSINTERNONVIDE) at (21,3) {$S_s \cap S_c \neq \emptyset$};
	
     \draw (NONSEG) -- (ENSINTERVIDE);
     \draw (NONSEG) -- (ENSINTERNONVIDE);

     \node[ok] (ENSINTVIDEUNIONN) at (6,-2) {$S_s \cup S_c = \mathds{N}$};
     \node[ok] (ENSINTVIDESUBSETN) at (12,-2) {$S_s \cup S_c \subset \mathds{N}$};

     \node[ok] (ENSINTNONVIDEUNIONN) at (18,-2) {$S_s \cup S_c = \mathds{N}$};
     \node[open] (ENSINTNONVIDEUNIONAUTRE) at (24,-2) {$S_s \cup S_c \neq \mathds{N}$};
	
     \draw (ENSINTERVIDE) -- (ENSINTVIDEUNIONN);
     \draw (ENSINTERVIDE) -- (ENSINTVIDESUBSETN);

     \draw (ENSINTERNONVIDE) -- (ENSINTNONVIDEUNIONN);
     \draw (ENSINTERNONVIDE) -- (ENSINTNONVIDEUNIONAUTRE);





    

    
    
    
  \end{footnotesize}
\end{tikzpicture}

    \caption{Leaves of the tree represent the general cases. Rounded corners green boxes corresponds to cases for which results are presented in this Section and in Section \ref{sec:segments}. Dashed boxes are cases not covered within this report.}
    \label{fig:generalcases}
\end{figure}

General cases correspond to all cases for which both $S_c$ and $S_s$ are non empty sets and none of both sets are equal to $\mathds{N}$. 
We classify all possible cases according to the tree represented on Figure \ref{fig:generalcases}.

The case $S_C = S_S$ composed of consecutive integers will be considered in section \ref{sec:segments}.
In the present Section we consider the cases for which $S_C \neq S_S$.
\begin{itemize}
    \item if $S_c \cap S_s = \emptyset$ and $S_c \cup S_s \subset \mathds{N}$ then the order of the graph is non-increasing.
    \item if $S_c \cap S_s \neq \emptyset$ and $S_c \cup S_s = \mathds{N}$ then the order of the graph is non-decreasing.
    \item If  $S_c \cap S_s = \emptyset$ and $S_s \cup S_c = \mathds{N}$, then $|V_t| = |V_0|$, the order of the graph is constant. 
\end{itemize}
These three cases are carefully analysed below.
In a first time the two sets are assumed disjoints.
On the second time they are assumed to cover the whole set of natural integer numbers.
Finally a consequence of the two first results is given assuming the two sets make a partition of $\mathds{N}$, which means they are disjoints and their union is $\mathds{N}$.

\subsubsection{\texorpdfstring{$S_s \cap S_c = \emptyset$}{Sc cap Ss = 0}}

\begin{theorem}
{Disjoint sets}
\label{theo:disjoint}
\\
Let $t\geqslant0$ and $G_t = (V_t,E_t)$ a graph and $S_s$ and $S_c$ two sets of positive integers. If $S_s\cap S_c = \emptyset$, then the series $(|V_t|)_{t \geqslant 0}$ is decreasing.
\end{theorem}

\noindent {\bf Proof}

Let $(G_t=(V_t,E_t))_{t \geqslant 0}$ a graph and $S_s$ and $S_c$ two disjoint sets.
To prove the result, it is enough to prove that for all $t \geqslant 0$, $|V_{t+1}| \leqslant |V_t|$.

Let $t \geqslant 0$ and $u \in V_t$. 
As $S_s\cap S_c = \emptyset$, nodes in $V_t$ can be divided in three groups:
\begin{itemize}
\item $V_s = \{u, \mbox{deg}(u) \in S_s~\mbox{and deg(u)} \notin S_c\}$, 
\item $V_c = \{u, \mbox{deg}(u) \in S_c~\mbox{and deg(u)} \notin S_s\}$
\item $V_{others} = \{u, \mbox{deg}(u) \notin S_c~\mbox{and deg(u)} \notin S_s\}$
\end{itemize}

From the rules, each node in $V_s$ is conserved, each node in $V_c$ is removed but generates a new node and each node in $V_{others}$ is removed, thus: 

$|V_{t+1}| = |V_s|+|V_c| \leq |V_s|+|V_c|+|V_{others}| = |V_t|$ 
\qed


\subsubsection{\texorpdfstring{$S_s \cup S_c = \mathds{N}$}{Ss cup Sc = N}}

\begin{theorem}{Union set}\label{theo:union}
\\
Let $S_s$ and $S_c$ subsets of $\mathds{N}$. If $S_s \cup S_c = \mathds{N}$, then the series $(|V_t|)_{t \geqslant 0}$ is increasing.
\end{theorem}

\noindent {\bf Proof}

Let $(G_t=(V_t,E_t))_{t \geqslant 0}$ a graph and $S_s$ and $S_c$ two non disjoint sets.
To prove the result, it is sufficient to prove that for all $t \geqslant 0$, $|V_{t+1}| \geqslant |V_t|$.

As for previous case, let $t \geqslant 0$ and $u \in V_t$. 
As $S_s\cup S_c = \mathds{N}$, nodes in $V_t$ can be divided in three groups:

\begin{itemize}
\item $V_s = \{u, \mbox{deg}(u) \in S_s~\mbox{and deg(u)} \notin S_c\}$, 
\item $V_c = \{u, \mbox{deg}(u) \in S_c~\mbox{and deg(u)} \notin S_s\}$
\item $V_{others} = \{u, \mbox{deg}(u) \in S_c~\mbox{and deg(u)} \in S_s\}$
\end{itemize}

From the rules it comes: each node in $V_s$ is conserved, each node in $V_c$ is removed but generates a new node and each node in $V_{others}$ is conserved and generates a new node, thus: 

$|V_{t+1}| = |V_s|+|V_c|+2|V_{others}| \geq |V_s|+|V_c|+|V_{others}| = |V_t|$ 
\qed


\subsubsection{Partition sets}

In this section, $S_S$ and $S_C$ are considered to be a partition of $\mathds{N}$. This means $S_S \cap S_C = \emptyset$ and $S_S \cup S_C = \mathds{N}$. From theorems \ref{theo:disjoint} and \ref{theo:union}, one can say that for every graph $G_t=(V_t,E_t)$, the series $(|V_t|)_{t_ \geqslant 0}$ remains steady.
Two cases rises from that situations:

\begin{itemize}
    \item $S_S=\mathds{N}$ \textbf{and} $S_C=\emptyset$: in that case the graph is constant ($\forall t, G_t = G_0$).
    \item $S_S=\emptyset$ \textbf{and} $S_C=\mathds{N}$: the series of static graphs $(G_t)_{t \geqslant 0}$ is a series of independent random geometric graph with a constant number of nodes ($\forall t, n_t=n_0$).
\end{itemize}

\section{Segments} 
\label{sec:segments}

This section focuses on the case $S_C=S_S=S$ where $S$ is a segment (i.e., an interval of consecutive integers).

\subsection{Model and conjecture}


In this section parameters $S_S$ and $S_C$ are limited to equal sets of consecutive integers. 
Both sets are such that $S_S=S_C=[m,M]$ (called segments), where $m, M \in {\mathds{N}}^2$ and referred to as $S$ in the following.
The evolution of graph order for different values of parameters $m$ and $M$ is investigated.
Some statements and properties are theoretically and experimentally proved for the special case $S=\{0\}$.
A relationship between graph order at a step $t+1$ and at step $t$ and an upper bound for $n_t$ ($t > 0$) are given.
Then, a theoretical analysis of the general case is provided, and a new concept named sustainable interval is introduced. 
In the last part of this section, vertices nervousness of graphs is studied through experimentation.
It is shown to be equal in average to $\frac{2}{3}$.
The reason behind this particular value will be explained in this last part.



\subsection{\texorpdfstring{$S=\{0\}$}{S=0}}

The case $S_S=S_C=S=\{0\}$ is considered in this section.
The seed graph, $G_0$, is supposed to be a random geometric graph whose order is arbitrarily chosen.
We first derive a bound on the graph order. 
Then the transition from step 0 to step 1 is studied.
Finally, the mean value of graph order is estimated.
An approximation for small values of the distance threshold $d$ is provided.

\begin{theorem}{(Bounded graph order)}
\\
Let $S=\{0\}$, $d \in ]0; \frac{1}{2}[$ and $G_0=(V_0, E_0)$ such that there exists at least one node $u \in V_0$ being isolated (i.e., $\mbox{deg}(u)=0$), then, for all $t > 0$, $n_t \leqslant \frac{8}{\pi d^2}$
\end{theorem}

\noindent {\bf Proof}

The unit torus and $0 < d < \frac{1}{2}$ are considered.
Let $S=\{0\}$ and $G_0=(V_0,E_0)$.
Let $t > 0$ such that $G_t$ is not empty. 
The size of $V_t$ is maximized as soon as there is not enough free space on the torus for adding a new isolated vertex. 
If we consider an empty torus, a vertex can be put anywhere. 
The area covered by this node is equal to $\pi d^2$.
For the rest of the proof, let consider, for each vertex $u$ on the torus, the disk of radius $d/2$ and center $u$ and referred it as $D(u)$.
From this, the condition for two nodes $u$ and $v$ to be non connected is: 
$D(u)\cap D(v) = \emptyset$.
After the addition of the first vertex $u_1$, a second vertex $u_2$ can be added to the torus if it satisfies $D(u_1) \cap D(u_2) = \emptyset$.
The area occupied by the two disks is then $2\times \pi (d/2)^2$, and the remaining free area is thus $1-2\times \pi (d/2)^2$.
Assuming $N$ vertices with non intersecting disks are already present in the torus, the free area is then $1-N\pi \frac{d^2}{4}$.

This quantity has to be positive so $N$ must be lower than or equal to $\frac{4}{\pi d^2}$.
This last quantity is an upper bound for the number of isolated vertices.
However, the rules says that isolated vertices duplicate. 
Thus, the number of vertices at one step $t$ can not exceed twice the upper bound of isolated vertices.
Hence, for all $t>0$, $n_t \leqslant \frac{8}{\pi d^2}$.\qed \\

This theorem provides an upper bound for graph order for $S=\{0\}$, ensuring graph order cannot exceed a certain value.
However, this does not provide any information about graph order evolution, which is the purpose of the two following theorems.

\begin{theorem}\label{theo:RGGOrderS1}
{(Expected graph order at step 1)}
\\
Let $S=\{0\}$, $d>0$ and $G_0=(V_0, E_0)$ be a random geometric graph of order $n_0$, then $\frac{n_1}{2} \sim B(n_0, (1-p(d))^{n_0 - 1})$, where $p(d)$ is the area of a circle of radius $d$ on the torus. An expectation value for $n_1$ is therefore $2n_0\cdot(1-p(d))^{n_0-1}$.
\end{theorem}

\noindent {\bf Proof}

For sake of clarity, in the remaining part of this proof, $p(d)$ will be referred to as $p$.
At the very first step $t=0$, $G_t=G_0$ is a random geometric graph and its nodes are uniformly distributed over the unit torus.
Let $(u, v)\in {V_0}^2$.
For a fixed threshold $d$, let consider the probability that $u$ and $v$ are connected.
$v$ is connected to $u$ if and only if $\mbox{dist}(u,v) \leqslant d$.
It means $v$ is in the disk of center $u$ and radius $d$.
If we denote by $X(u,v)$ the event ``$u$ and $v$ are connected", the wanted probability is the ratio of the area of the surface of a disk of radius $d$ over the area of the unit torus.
For all $(u,v)\in {V_0}^2$, $X(u,v) \sim B(p)$.

Let's study the degree distribution of a node $u\in V_0$.
As the position of every point is independent one from the others, variables $X(u,x)$ are independent for all $x\neq u$.
More over the degree of $u$ is the number of connections $u$ has to other nodes :
\[
\mbox{deg}(u)=\sum_{u\neq v}X(u,v)
\]
All variables $X(u,v)$ being independent for all $v\neq u$,
$\mbox{deg}(u) \sim B(|V_0|-1, p)$ as a sum of $|V_0|-1=n_0-1$ independent Bernoulli variables of same parameter p.
Let's consider $Y_0(u)$ the event ``$u$ is conserved at step 1" ($Y_0(u)=1$ if and only if $u$ is conserved and 0 otherwise), then, as $S=\{0\}$:
\[
P(Y_0(u)=1)=P(\mbox{deg}(u)=0)=(1-p)^{n_0-1}
\]
This means ${Y_0(u)} \sim B((1-p)^{n_0-1}) $.
Thus the number of conserved vertices at step 1 is: \[
Y_0=\sum_{u\in V_0}{Y_0(u)}
\]
As all points have an independent position, $\frac{n_1}{2} = Y_0 \sim B(n_0, (1-p)^{n_0-1})$.\qed

\begin{theorem}\label{theo:RGGOrder}
{(Expected value of graph order)}
\\
Let $S=\{0\}$, $d>0$ and $G_0=(V_0, E_0)$ such that there exists at least one node $u \in V_0$ being isolated (i.e., $\mbox{deg}(u)=0$), then
either the graph becomes empty, or the average number of conserved nodes is $l(d) = 1-\frac{\log{(\frac{\sqrt{1+4\alpha}-1}{2}})}{\log{\alpha}}$ with $\alpha=\frac{1}{1-p}$ and $p=p(d)$.

\end{theorem}


\noindent {\bf Proof}

Let $t \geqslant 1$. Two cases are to be discussed:  the case of conserved vertices from step $t-1$ to step $t$ ($V_t \cap V_{t-1}$) and the case of created nodes at step $t$ ($V_t - V_{t-1}$).
As the number of created nodes is the same as the number of conserved nodes from $t-1$ to $t$, we set $c_t=|V_t \cap V_{t-1}|=|V_t - V_{t-1}|$.

First let's study the number of conserved vertices from step $t$ to step $t+1$ among those conserved from step $t-1$ to step $t$.
$c_{t+1}^{\mbox{\footnotesize conserved}}$ denotes this number. 
Let $u\in V_t \cap V_{t-1}$. The probability for $u$ to be conserved is the probability that its degree to created nodes remains equal to 0.
\[
\mbox{deg}(u)=\sum_{v\in V_t-V_{t-1}}X(u,v)
\]
Let $v \in V_t-V_{t-1}$.
As in the previous section, $X(u,v) \sim B(p)$ and $\mbox{deg}(u) \sim B(c_t, p)$ as a sum of independent Bernoulli variables of same parameter $p$.
$Y_t(u)$ denotes the event ``$u$ is conserved at step $t+1$".
The probability that $u$ survives is $P(Y_t(u)=1)=P(\mbox{deg}(u)=0)=(1-p)^{c_t}$, thus: $Y_t(u) \sim B((1-p)^{c_t})$.
Therefore, the number of conserved vertices at step $t+1$ among those conserved at step $t$ is: 
\[
c_{t+1}^{\mbox{\footnotesize conserved}}=\sum_{u\in V_t \cap V_{t-1}}Y_t(u)
\]
As the position of created nodes are independent from themselves and from conserved vertices, $Y_t(u)$ are independent for all $u\in V_t \cap V_{t-1}$, $c_{t+1}^{\mbox{\footnotesize conserved}} \sim B(c_t, (1-p)^{c_t})$.

Let's study the number of conserved vertices among created nodes.
$c_{t+1}^{\mbox{\footnotesize created}}$ denotes this number. 
Let $u \in V_t-V_{t-1}$. To study the degree of $u$, two cases must be studied. 
The first one is the number of connections between $u$ and all other created nodes (denoted as $\mbox{deg}^C(u)$). The second one is the number of connections to already present nodes (denoted as $\mbox{deg}^S(u)$).
$\mbox{deg}^C(u)$ and $\mbox{deg}^S(u)$ can be obtained using the following formulas:

\begin{align*}
    \mbox{deg}^C(u) &= \sum_{v\in V_t - V_{t-1}, u \neq v }X(u,v) \\
    \mbox{deg}^S(u) &= \sum_{v \in V_t \cap V_{t-1}}X(u,v)
\end{align*}

As the position of created points on the torus are independent one from the others, $\mbox{deg}^C(u)$ is a sum of independent Bernoulli variables and therefore,  
$\mbox{deg}^C(u) \sim {B(c_t -1, p)}$. For $\mbox{deg}^S(u)$, connections between a created node and an already present node are not independent from each other: knowing $u$ is connected to an already present node means it is close to it and as other conserved nodes are farther than $d$, it implies that $\mbox{deg}^S(u)$ is not a sum of independent Bernoulli variables.
However, as a first approximation, this quantity will be considered as a sum of independent Bernoulli variables.

Thus, the computation of the expectation of $c_{t+1} = c_{t+1}^{\mbox{\footnotesize conserved}}+c_{t+1}^{\mbox{\footnotesize created}}$ gives:
\[
c_{t+1}=c_t(1-p)^{c_t}+c_t(1-p)^{2c_t-1}
\]
By looking for a limit to this series gives $l \geqslant 0$ satisfying:
\[
l=l(1-p)^l+l(1-p)^{2l-1}
\]
Solving this equation gives $l=0$ or :
\begin{eqnarray*}
    l = 1 -  \frac{\log{\left({\frac{\sqrt{1+4\alpha}-1}{2}}\right)}}{\log{\alpha}} & & \textnormal{with $\alpha=\frac{1}{1-p}$}
\end{eqnarray*}
\qed 

\begin{figure}[h!]
  \begin{center}
    \begin{tikzpicture}
      \begin{axis}[
          width=\linewidth, 
          grid=major, 
          grid style={dashed,gray!30}, 
          xlabel=$c_t$, 
          ylabel=Expected value,
          title style={yshift=0.5ex, align=left},
          title={Relationship between the average value of $c_t$ and\\the expected value mentioned in theorem \ref{theo:RGGOrder}.},
          xmin = 0, xmax = 155000,
          ymin = 0, ymax = 155000
        ]
        \addplot[color=blue!50!cyan,only marks]
        table[x=Expected,y=Survivors,col sep=comma] {NodeCountExpected.csv};
        \addplot[color=black,smooth,dashed,tension=0.7,very thick] coordinates{(0,0)(155000,155000)};
        \legend{$c_t=f($Expected average value$)$, Identity ($y=x$)}
      \end{axis}
    \end{tikzpicture}
    \caption{Relationship between the average value of $c_t$ and the expected value. Each point correspond to a single threshold $d$. $d$ is ranging from 0.001 to 0.01 with a step of 0.0005 and from 0.01 to 0.2 with a step of 0.005}
    \label{fig:expected_relationship_isolated}
  \end{center}
\end{figure}
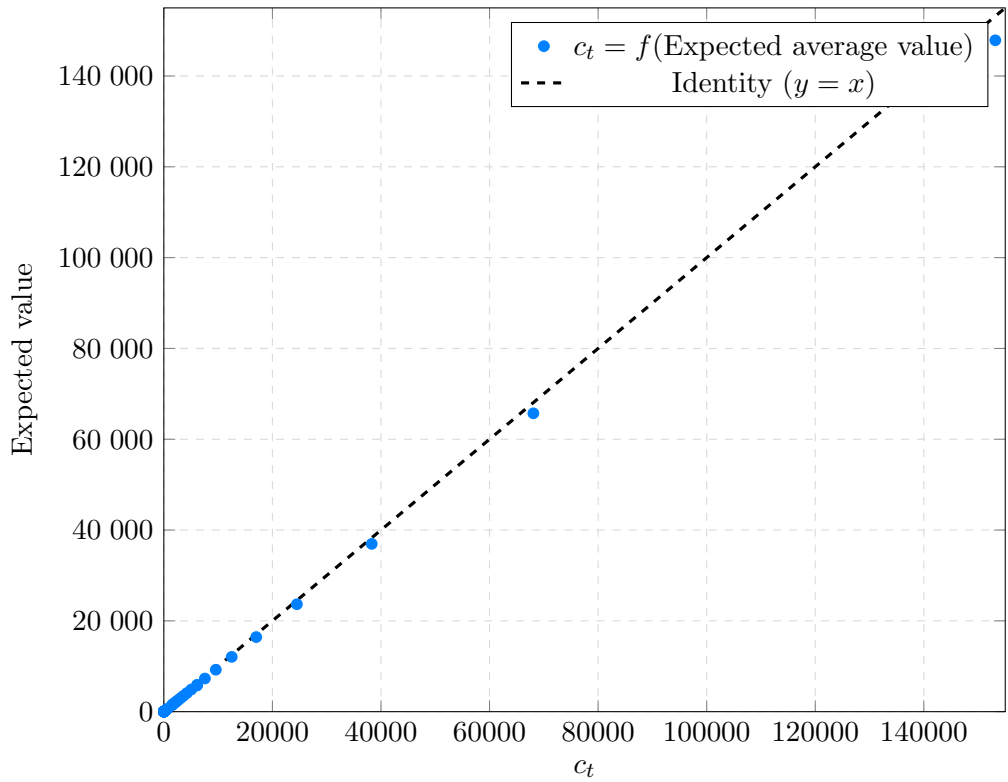

Experiments have been run to see if this relationship holds.
The results are summarized on figure \ref{fig:expected_relationship_isolated}.
This figure shows the accuracy of the expectation of graph order given in theorem \ref{theo:RGGOrder}.
Indeed, the blue curve, close to the dashed line, highlights that theoretical expectation and experimental results are close to be equal.
This result being proved leads to an approximation for small values of threshold $d$.

\begin{corollary}
Let $d > 0$ and $l(d)$ as defined in the previous theorem (\ref{theo:RGGOrder}). Then for small values of $d$:
\[l(d) \sim -\frac{\log{\left(\frac{\sqrt{5}-1}{2}\right)}}{\pi d^2} = \frac{\log{\phi}}{\pi d^2}\]
where $\phi$ is the golden ratio $\left(\frac{1+\sqrt{5}}{2}\right)$.
\end{corollary}

\noindent{\textbf{Proof}}
Let $d > 0$ be small. Thus, applying Taylor expansion gives $\frac{1}{1-\pi d^2} \sim 1 + \pi d^2$ and $\log{\left(\frac{1}{1-\pi d^2}\right)} \sim \pi d^2$.
The numerator comes from $4 \cdot \frac{1}{1-\pi d^2} \simeq 4$.
The golden ratio is obtained using operations on $\log$ and by noticing that $\frac{2}{\sqrt{5}-1} = \frac{2(\sqrt{5}+1)}{4} = \phi$, the golden ratio.
Combining these results leads to the statement of the corollary.
\qed\\

It is therefore possible to state that, in the case where $S=\{0\}$, it is possible to theoretically get an expectation of graph order as well as to get an upper bound for graph order depending on parameter $d$.


\subsection{The general case}

Now the focus is on $S=[m, M]$ for every $m$ and $M$ integers.
The goal is to provide a tool aiming at stating, for given parameters $m, M$ and $d$, whether the graph is likely to be sustainable or not.
This part mainly focuses on a simpler model.
This model is studied as it helps understanding the evolution of graph order.


\subsubsection{Study of graph evolution}

In this Section we aim at estimating the evolution of the graph order during graph dynamics.
However, in the D3G3 model, between two time steps, non-conserved nodes are removed from the graph and conserved nodes are located at the same position, which entails a remanent graph.
This remanent graph induces a structure influencing the computation of graph order.
More precisely, nodes that are about to be removed connected to conserved ones interfere in the probability that conserved nodes at time $t$ are still conserved at time $t+1$.
This is linked to computing the degree of the neighbors of a node $u$ knowing the degree of node $u$.
To our knowledge, this is a difficult question.
For that purpose, a relaxed version of the D3G3  model is considered enabling analytical study of this evolution.
In this model, conserved nodes are moved (i.e., their position are changed) such that obtained graph is a new random geometric graph at each step.
We call this model "the redistributed model".
This will help us proving the following theorem:

\begin{theorem}\label{theo:FRGGP_law}
Let $G=(G_t)$ be a dynamic graph obtained with the redistributed model, then at every step $t$, $\frac{n_{t+1}}{2} \sim B(n_t, p(S, d, n_t))$, where $p(S, d, n_t)$ is the probability that a node is conserved between step $t$ and $t+1$:
\[
p(S, d, n_t) = \sum_{k = m}^M{n_t - 1 \choose k}p^k(1-p)^{n_t - 1 - k}
\]
Here, $p(d)$ refers to the probability for two different nodes to be connected (i.e., the probability that the distance between them is lower than or equal to $d$), which is, for $d \leqslant \frac{1}{2}$, $\pi d^2$.
\end{theorem}

\noindent{\bf Proof}

{In the redistributed model, at time step $t$ a RGG ($G_t$) is built. 
If the graph order at time $t$ is equal to $n_t$, the graph order at $t+1$ is equal to twice the number of surviving nodes at time $t$.}
As every node has an independent position {in the torus}, this probability is the same for all nodes. 
Let's denote it $p(S, d, n_t)$.
Let $u \in V_t$.
Then:
\begin{align}
p(S, d, n_t) &= P(\mbox{deg}(u) \in S) = \sum_{k = m}^MP(\mbox{deg}(u) = k) \\
p(S, d, n_t) &= \sum_{k = m}^M{n_t - 1 \choose k}p^k(1-p)^{n_t - 1 - k} \label{eq:probaSurviving}
\end{align}
Assuming one node is a conserved node does not affect the probability of conservation for other nodes. 
The number of conserved nodes can be computed summing independent Bernoulli's events of parameter $p(S, d, n_t)$.
This gives $\frac{n_{t+1}}{2}$ follows a binomial distribution of parameter $n_t$ and $p(S, d, n_t)$.
\qed \\

Computing expectation for a binomial distribution leads to an expectation for $n_{t+1}$ knowing $n_t$.
Indeed, this expectation is $2n_tp(S, d, n_t)$.
For a fixed set $S$, this provides a relationship between $n_t$ and $n_{t+1}$:
\begin{definition}
\underline{Expectation of graph order}:\\
Let $m$, $M$ and $d$ be parameters for the redistributed model.
Let $G=(G_t)$ be an obtained graph with such parameters.
Then, the expectation of graph order at step $t+1$ ($n_{t+1}$) knowing graph order at step $t$ ($n_t$) is $f_{S,d}(n_t)$ and satisfies $n_{t+1} = f_{S,d}(n_t) = 2n_t p(S,d,n_t)$, and then:
\begin{equation}
\forall n \in \mathds{N}, f_{S,d}(n) = 2n p(S,d,n)
\end{equation}
\end{definition}

This quantity is referred to as the relationship in the sequel.
Studying the relation for every value of $m$, $M$ and $d$ turns out to be a difficult problem.
However some results may be conjectured.
A first conjecture concerns the variations of the relationship:
\begin{conjecture}\label{cj:increase_decrease}
Let $m, M$ and $d$ be parameters of the model. Let $S=[m, M]$ and $f_{S,d}$ the relationship as defined above. Then there exists $n_* \in \mathds{N}$ such that $f_{S,d}$ is increasing on $[0, n_*]$ and decreasing on $[n_*+1, +\infty[$.
\end{conjecture}
This conjecture is difficult to prove due to the sum involved in the computation of $f_{S,d}$.
However, it is not necessary to study the relationship for all integers.
It is possible to perform the study on a limited interval.
This is the purpose of theorem \ref{theo:increase_decrease} (below).
But before proving this theorem, it is necessary to provide another formulae computing variations of $f_{S,d}$:
\begin{lemma}
\label{lm:variation}
Let $m$, $M$ and $d$ be parameters of the model.
Let $\Delta f_{S,d}$ defined as the variation of $f_{S,d}$: for $n \in \mathds{N}, \Delta f_{S,d}(n)= f_{S,d}(n+1) - f_{S,d}(n)$.
Then:
\[
\forall n \in \mathds{N}, \Delta f_{S,d}(n) = 2\sum_{k=m}^M {(k+1){n \choose k}p^k(1-p)^{n-1-k}}\left(1-\frac{n+1}{k+1}p\right)
\]
\end{lemma}

\noindent \textbf{Proof}:
Let $m$, $M$ and $d$ be parameters of the model.
Let $n$ a be non-negative integer.
This proof only focuses on the terms of the sum of $\Delta f_{S,d}$:
\begin{align*}
\Delta f_{S,d}(n) &= 2\left(\sum_{k=m}^M(n+1){n \choose k}p^k(1-p)^{n-k}-n{n-1 \choose k}p^k(1-p)^{n-1-k}\right) \\
&= 2\sum_{k=m}^M{p^k(1-p)^{n-1-k}\left((n+1){n \choose k}(1-p) - {n-1 \choose k}\right)}
\end{align*}
Let $k \in \mathds{N}$ such that $m \leqslant k \leqslant M$.
Every term of the sum of $\Delta f_{S,d}$ can be expressed as follow only using results on binomial coefficients:
\begin{align*}
(n+1){n \choose k}(1-p) - n{n-1 \choose k} &= (k+1){n+1 \choose k+1}(1-p) - (k+1){n \choose k+1} \\
&= (k+1)\left({n+1 \choose k+1}-p{n+1 \choose k+1} - {n \choose k+1}\right) \\
&= (k+1)\left({n \choose k} - p{n+1 \choose k+1}\right) \\
&= (k+1)\left({n \choose k} - p\frac{n+1}{k+1}{n \choose k}\right) \\
&= (k+1){n \choose k}\left(1-p\frac{n+1}{k+1}\right) \\
\end{align*}
\vspace{-0.1in}
This last equality leads to the following form of $\Delta f_{S,d}$:
\begin{equation*}
\Delta f_{S,d}(n) = 2\sum_{k=m}^M {(k+1){n \choose k}p^k(1-p)^{n-1-k}}\left(1-\frac{n+1}{k+1}p\right)
\end{equation*}
\qed

It is now possible to state the following theorem about variations of $f_{S,d}$:
\begin{theorem}
\label{theo:increase_decrease}
Let $m, M$ and $d$ be the parameters of the model. Let $S=[m, M]$ and $f_{S,d}$ the relationship as defined above. Let $p=p(d)$ be the probability for two different nodes to be connected. Then, $f_{S,d}$ is increasing between 0 and $\frac{m+1}{p}-1$ and decreasing from $\frac{M+1}{p}-1$ to infinity.
\end{theorem}

\noindent \textbf{Proof}:
The goal is to prove that $\Delta f_{S,d}(n)$ is positive for $n < \frac{m+1}{p}-1$ and negative for $n > \frac{M+1}{p}-1$.
To understand this, $\Delta f_{S,d}(n)$ can be rewritten as follow (lemma \ref{lm:variation}):
\[
\forall n \in \mathds{N}, \Delta f_{S,d}(n) = 2\sum_{k=m}^M {(k+1){n \choose k}p^k(1-p)^{n-1-k}}\left(1-\frac{n+1}{k+1}p\right)
\]
It is sufficient to notice that, for all $k \in S$, the sign of every single term of the sum is the sign of $\left(1-\frac{n+1}{k+1}p\right)$.
For fixed $k$, the term is positive if and only if $n$ is lower than $\frac{k+1}{p}-1$.
As this last term is an increasing function of $k$,
all terms of the sum are therefore positive if $n$ is lower than $\frac{m+1}{p}-1$ and negative if $n$ is greater than $\frac{M+1}{p}-1$.
Hence, the relationship is increasing from 0 to $\frac{m+1}{p}-1$ and decreasing from $\frac{M+1}{p}-1$ to infinity.
\qed\\

\noindent Thanks to theorem \ref{theo:increase_decrease}, conjecture \ref{cj:increase_decrease} is proved for intervals $[0,x_m]$ and $[x_M, \infty[$ $x_m=\frac{m+1}{p}-1$ and $x_M=\frac{M+1}{p}-1$. 
At this stage, quantifying more precisely the evolution of the graph order is not achievable. 
However, a study of the fixed points of $f_{S,d}$ enables to draw some conclusion about generated graphs sustainability.



\subsubsection{Graph evolution and sustainability}

First note that knowing the variations of $f_{S,d}$ is not enough to deal with graphs sustainability.
Indeed, as claimed by the following theorem, big graphs are not sustainable. 
\begin{theorem}
\label{theo:big_graphs_collapse}
\underline{Non-sustainability of big graphs}:\\
Let $m$, $M$ and $d$ be parameters of the model.
Let $f_{S,d}$ be the relationship.
Then, there exists $N>0$ such that for all $n > N, f_{S,d}(n) < 1$.
\end{theorem}

\noindent \textbf{Proof}:
For this proof, it is sufficient to prove that $f_{S,d}(n) \to 0$ when $n \to +\infty$.
Let $n$ such that $n \geqslant 2M+1$.
In this situation, for all $k \leqslant M$, binomial coefficient ${n-1 \choose k} \leqslant {n-1 \choose M}$.
Moreover, as $(1-p) < 1$, $x \longmapsto (1-p)^x$ is decreasing.
Therefore, for all $k \leqslant M$, $(1-p)^{n-1-k} \leqslant (1-p)^{n-1-M}$.
It is thus possible to get the following inequality for all $k \leqslant M$:
\begin{equation*}
{n-1 \choose k}(1-p)^{n-1-k}p^k \leqslant {n-1 \choose M}(1-p)^{n-1-M}p^k
\end{equation*}
Noticing $p < 1$ and $f_{S,d}$ is a sum of $M-m+1$ elements, $f_{S,d}(n)$ can be bounded as follow
\begin{equation*}
f_{S,d}(n) \leqslant 2n(M-m+1)\left({n-1 \choose M}(1-p)^{n-1-M}\right)
\end{equation*}
As $M$ is fixed, the binomial coefficient ${n-1 \choose M}$ is equivalent to a polynomial of degree $M$ as $n$ grows to infinity:
\begin{equation*}
{n-1 \choose M} \sim \frac{n^M}{M!}
\end{equation*}
Therefore, $f_{S,d}(n)$ is equivalent to the product of a polynomial and an exponential function converging towards 0.
This implies $f_{S,d}(n)$ converges towards 0 as $n$ tends to infinity.
\qed \\
This theorem says that there always exists a graph order limit such that graphs whose order are greater than this limit are likely to become empty.
Therefore, it is not possible to obtain sustainable graphs with a large amount of nodes.

A new mathematical concept is now introduced aiming at classifying parameters into three classes.
This concept is referred to fixed point and is defined as follow:
\begin{definition}
\underline{Fixed Point}:\\
Let $m$, $M$ and $d$ be parameters of the model.
A fixed point for the relationship $f_{S,d}$ is an non-negative integer $n$ such that:
$$\left\{ 
  \begin{array}{ c l }
    & f_{S,d}(n) \leqslant n ~\mbox{and} ~f_{S,d}(n+1) > n+1 \\
    \text{or} & f_{S,d}(n) \geqslant n ~\mbox{and} ~f_{S,d}(n+1) < n+1
  \end{array}
\right.$$
\end{definition}
Such fixed points characterize variation of graph order.
Indeed, graph of order $n$ for $n$ taken between two consecutive fixed points is either always decreasing or increasing.
From experiment performed on the redistributed model as well as on D3G3, three different cases appear and are conjectured as follow:
\begin{conjecture}
\label{cj:three_fixed_point}
For all $m$, $M$ and $d$ being parameters of the model, the relationship $f_{S,d}$ has either one, two or three fixed points.
\end{conjecture}
This conjecture is the main tool aiming at studying sustainability in the segment case.
Indeed, in the three different cases, it is possible to answer whether a given set of parameters is sustainable or not.
However, their is no characterisation about parameters value that may help founding which case parameters lead to.
The only one claim that can be made is that $d$ does have an influence on this case.

The conjecture \ref{cj:three_fixed_point} is assumed in this subsection.
This section aims at stating about sustainability in the three different cases.
This is illustrated by a description of the behavior of the relationship $f_{S,d}$ in every case.
 
\subsubsubsection{One fixed point}

\begin{figure}[!h]
    \centering
    \includegraphics[width=0.9\textwidth]{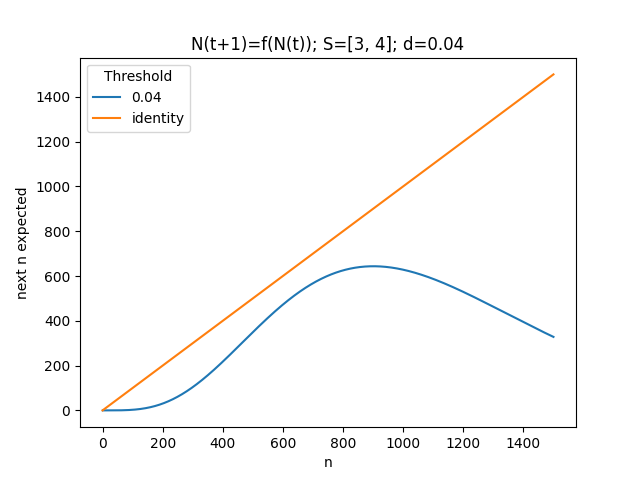}
     \caption{One fixed point}
     \label{sub:One_fixed_point}
\end{figure}

First let's consider the case where the relationship has only one fixed point.
When it has only one fixed point, this point is $0$.
This comes from $f_{S,d}(0) = 0$.
Moreover, for all $n$, $f_{S,d}(n) < n$.
As for a snapshot graph of order $n_t$ at step $t$, $f_{S,d}(n_t)$ gives the expectation value of $n_{t+1}$ at step $t+1$.
Graph orders of generated graphs are decreasing in average.
Graphs obtained in this case are therefore not sustainable.
This is illustrated by Figure \ref{sub:One_fixed_point}.
 
\subsubsubsection{Two fixed points}

\begin{figure}[!h]
    \centering
    \includegraphics[width=0.9\textwidth]{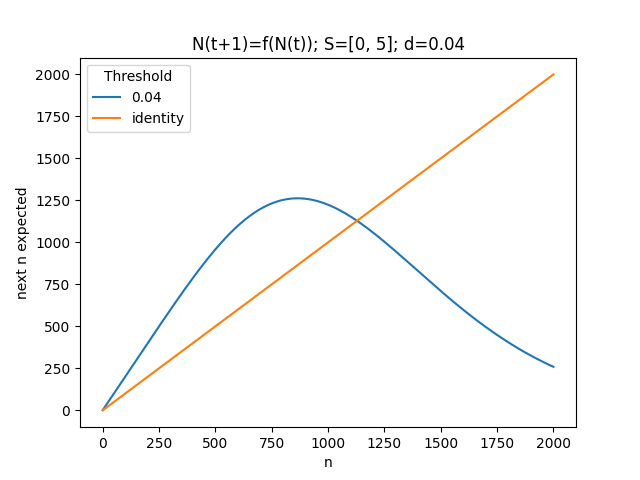}
     \caption{Two fixed points}
     \label{sub:Two_fixed_points}
\end{figure}

For the two fixed points case, 0 is also a fixed point.
The argument is also because $f_{S,d}(0)=0$.
The other one is greater than zero.
The case where $f_{S,d}$ has two fixed points is assumed to happen if and only if $0 \in S$ and is stated in the following conjecture.
\begin{conjecture}{(Characterisation of the two fixed points)}
\label{cj:caracterisation_of_two_fp}
The relationship $f_{S,d}$ has two fixed points if and only if $m=0$.
\end{conjecture}
An argument is that a snapshot graph with one vertex becomes empty if and only if $0 \notin S$, that is $m=0$.
Moreover, for $n=0$, the first term of the sum defining $f_{S,d}$ is equal to 1 so $f_{S,d}(0) = 2$.
Graphs generated in such configurations are therefore sustainable as long as their graph order does not exceed a limit.
This limit is a consequence of theorem \ref{theo:big_graphs_collapse}.
In this case, graphs whose order exceeds the limit are likely to become empty.
This is illustrated by figure \ref{sub:Two_fixed_points}.
 
\subsubsubsection{Three fixed points}

For the last case, the goal is to show that graph order is likely to remain bounded.
Deeply looking at this case raises the question of values of graph order for which the size is not too large and not too small so that it does not collapse.
{For that purpose we define an interval, called \textit{sustainable interval}, such that, if the graph order remains within that interval, this ensures the persistence of the graph.}
This sustainable interval is considered as a tool to study graph sustainability.
It concerns expectation of graph order evolution through time.
It says that if the image of the function $f_{S,d}$ for all integers within the interval does not exceed the upper bound, then the graph is likely not to collapse.
Let's define more precisely this concept:
\begin{definition}
\underline{sustainable interval}:\\
Let $m$, $M$ and $d$ be parameters of the model.
Let consider $f_{S,d}$ set such that it has three fixed points.
Let $N_m$ be the first positive fixed point and  $N'_m$ the smallest integer greater than $N_m$ such that $f_{S,d}(N'_m) \geqslant N_m$ and $f_{S,d}(N'_m + 1) < N_m$. 
The \textit{sustainable interval} associated to $m$, $M$ and $d$ is defined as the interval $[N_m, N'_m]$.
\end{definition}

\begin{figure}[!h]
    \centering
    \includegraphics[width=0.9\textwidth]{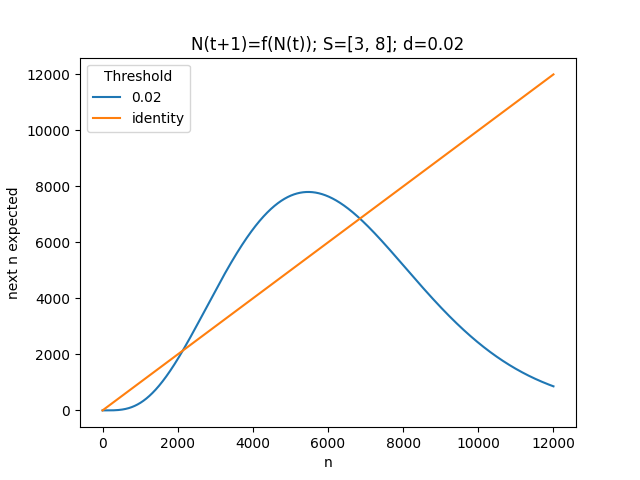}
     \caption{Three fixed points}
     \label{sub:Three_fixed_points}
\end{figure}

This definition is illustrated through figure \ref{sub:Three_fixed_points}.
Such an interval satisfies a property about the values $f_{S,d}$ takes when it is restricted to it:
\begin{theorem}
\underline{Sustainability in the sustainable interval}:\\
Let $m$, $M$ and $d$ be parameters of the model.
Let assume the relationship $f_{S,d}$ has three fixed points and that $[N_m, N'_m]$ is its associated sustainable interval.
Then, the relationship satisfies:
\begin{equation*}
\forall n \in [N_m, N'_m], f_{S,d}(n) \geqslant N_m
\end{equation*}
Moreover, if the relationship does not exceed $N'_m$, then the relationship satisfies:
\begin{equation*}
\forall n \in [N_m, N'_m], f_{S,d}(n) \in [N_m, N'_m]
\end{equation*}
\end{theorem}

\noindent \textbf{Proof:}
The first part of the theorem comes from definition of the sustainable interval.
\qed\\

\noindent Main interpretation of that theorem is graphs are sustainable in probability in the sustainable interval if and only if there are no values of $f_{S,d}$ that exceed the upper bound of the sustainable interval.\\

The following paragraphs provides arguments aiming at obtaining the sustainable interval.
They also provide arguments to check whether the relationship exceeds the upper bound of the interval.
The theorem \ref{theo:increase_decrease} clearly gives bounds to find out the maximum of the relationship $f_{S,d}$.
Three algorithms are sufficient to answer both questions:
an algorithm to compute the argument of the maximum of the relationship $f_{S,d}$, an algorithm to find its fixed point between 0 and the argument of the maximum and an algorithm to solve $f_{S,d}(n)=y$ for $n$ greater than the argument of the maximum and $y>0$ lower than or equal to the maximum.
In the following, these algorithms are first implemented.
It is then explained how to use them to answer questions about the sustainable interval.\\

\noindent \underline{\textbf{The argument maximum}}: To compute the argument maximum of the relationship, it is sufficient to study $f_{S,d}$ on the interval $[x_m, x_M]$ for $x_m$ and $x_M$ as defined above.
This is a consequence of theorem \ref{theo:increase_decrease}. Let's denote it $N_*$.\\

\noindent \underline{\textbf{The first positive fixed point}}: To find the fixed point of $f_{S,d}$ mentioned in the definition of the sustainable interval, it is sufficient to compute the argument maximum of it. The previous algorithm answers this question.
Then, as the relationship is increasing from 0 to $N_*$, it is sufficient to iterate and find an integer $n$ such that $f_{S,d}(n) \leqslant n$ and $f_{S,d}(n + 1) > n + 1$.\\

\noindent \underline{\textbf{The solution of the equation}}: For the last algorithm, the goal is to find an integer $n$ such that $n$ is greater than $N_*$ of $f_{S,d}$, $f_{S,d}(n) \geqslant y$ and $f_{S,d}(n) < y$, for a fixed $y$ which is assumed to be positive and lower than the maximum of $f_{S,d}$.\\

\noindent From these algorithms it is possible to implement algorithms stating the existence of the sustainable interval and its bounds.
For the existence or not of the sustainable interval, it is sufficient to check whether the maximum of the relationship is greater than its argument.
This comes from that sustainable interval exists if and only if there are values of the relationship that exceed their argument.
As the relationship is increasing from 0 to $f_{S,d}(N_*)$, then  sustainable interval exists if and only if $f_{S,d}(N_*) > N_*$.
For computing the sustainable interval boundaries, it is sufficient to know the value of the first fixed point $N_m$ (as it provides the lower bound) and to solve the equation $f_{S,d}(x)=N_m$ as finding the corresponding $x$ to this equation provides the upper bound ($N'_m$).
The existence of $N'_m$ is ensured by theorem \ref{theo:big_graphs_collapse}.

\subsection{Vertex nervousness}

The goal is to highlight a characterization aspect of the segment family using the vertex nervousness metric.
As edge nervousness will not be studied for that case, vertex nervousness will be referred to as nervousness in this section.
As in this particular configuration, survivors are the same as created nodes, it is possible to state particular results about the value of nervousness:
\begin{theorem}\label{theo:nervousness_segment}
Let $S$ be a segment set of non-negative integer and $d \in ]0, \frac{1}{2}[$. Let $G$ be a generated graph of order $n_t$ at step $t$ and number of survivor from step $t$ to step $t+1$ referred to as $s_t$. Then:
\[
    \mbox{VN}_t = \frac{n_t}{n_t+s_t}
\]
\end{theorem}

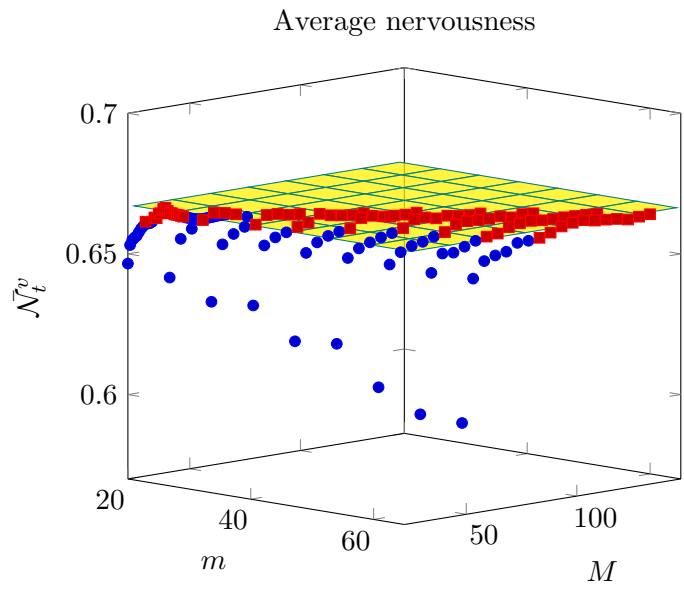
\begin{figure}
    \centering
\begin{tikzpicture}
\begin{axis}[
          view = {45}{10},
          width=0.7\linewidth, 
          xlabel=$m$, 
          ylabel=$M$,
          zlabel=$\bar{\mathcal{N}^v_t}$,
          legend style={at={(0.05.,.98)}}, 
          title style={yshift=0.5ex},
          title={Average nervousness},
          zmin=0.57, zmax=0.7,
          only marks
        ]
 
    \addplot3+
    [           
        samples= 9,
        on layer=background,
        z filter/.expression={z<0.66 ? z : nan}
    ]
    table [x=mS, y=MS, z=Nervousness,col sep=semicolon]{nervousness_mean.csv};
 
    \addplot3+
    [
        samples= 9,
        fill=red,
        on layer=foreground,
        z filter/.expression={z>0.66 ? z : nan}
    ]
    table [x=mS, y=MS, z=Nervousness,col sep=semicolon]{nervousness_mean.csv};

    \addplot3 [
    fill opacity=0.75,
    shader=faceted,
    domain=20:65,
    domain y = 25:145,
    samples = 8,
    samples y = 8,
    surf,
    fill=yellow,
    faceted color = teal,
    on layer=main] {0.6666666666};
 
\end{axis}
\end{tikzpicture}
    \caption{Mean value of nervousness got from experimentation.
    Points represent the average over 20 run and 30000 time steps for a single $m$ and $M$.
    The yellow surface is the plan of equation $z=\frac{2}{3}$.
    For all these parameters, $d$ is set to $0.05$.
    Red points represent nervousness of value greater than $\frac{2}{3}$.
    Blue points represent nervousness of value lower than $\frac{2}{3}$.
    }
    \label{fig:nervousess_mean}
\end{figure}

\noindent{\textbf{Proof}}
To prove this result, it is sufficient to notice that $n_{t+1} = 2s_t$, as $S_S=S_C$, which means the number of survivors is the same as the number of created nodes.
Thus, applying some basic result about set sizes and noticing that $s_t = |V_t \cap V_{t+1}|$, leads to:
\begin{align*}
    |V_t \cup V_{t+1}| &= n_t + n_{t+1} - s_t = n_t + s_t \\
    |V_t \triangle V_{t+1}| &= n_t + n_{t+1} - 2|V_t \cap V_{t+1}| = n_t
\end{align*}
It follows that vertex nervousness is well equal to $\frac{n_t}{n_t+s_t}$.
\qed\\
Result about the nervousness observed in generated graphs parameterized with a segment set $S$ is stated in the following conjecture:
\begin{conjecture}
Let $m, M \in \mathds{N}$. Let $S = [m, M]$ and $d > 0$ be parameters of the graph $G=(G_t)_{t \leqslant 0}$. Then the nervous of the graph is in average equal to $\frac{2}{3}$.
\end{conjecture}
Although this conjecture has not been proved theoretically, experimentation have been performed.
They all highlight this conjecture telling that the average nervousness of generated graphs is roughly equal to $\frac{2}{3}$.
Results of this experimentation are gathered on picture \ref{fig:nervousess_mean}.
A possible interpretation of this conjecture and performed experimentation relies on the result stated in theorem \ref{theo:nervousness_segment} and on results from last part.
Indeed, if vertex nervousness is close to $\frac{2}{3}$, it means $s_t\simeq\frac{n_t}{2}$.
Then, as $n_{t+1}=2s_t$, it comes $n_{t+1}\simeq n_t$, which means that graph order is close to a fixed point of the relationship $f_{S,d}$ mentioned in the previous section.

\section{Conclusion}
This paper shows our first investigations in the study of dynamic graph generators.
This work concerns a simple generator.
As a reminder, the model is parameterized through three variables: a connection threshold $d$ aiming at connecting all points closer than a distance $d$ and two sets $S_S$ and $S_C$ containing non-negative integers.
The first one aims at deciding whether a node is kept between two consecutive steps and the second one whether a node is at the origin of a new node at the very next step.
Several non-trivial properties are shown about the model.
All these properties concern products of the generator.
The generator, for a single configuration, produces a family of graphs and not a single graph.
Properties are therefore about the whole family of graphs the generator provides for a single configuration.
All these properties shown try to answer a single question.
This question concerns graph \textit{sustainability}. 
It is defined as the property, for a given graph obtained with a given seed graph and evolving rules, that the graph does not become empty after a finite number of steps.
Defining this concept for this model is not simple since the evolving rules are not deterministic.
It involves probabilistic computations and therefore questions about a possible threshold for which the graph is said to be sustained if the probability of the emptiness of the graph is greater than this threshold.
Here the focus has been made on two different metrics, graph order evolution and vertex nervousness, the second one being a renaming of the Jaccard distance metric.
Different values of the parameters have been studied, but it has not been possible to try them all as the amount of possible cases is far too big.
Cases for which properties have been shown are limit cases, the general case and a very specific case referred as "segments".
Limit cases have led to a first classification when at least one of the two parameter sets is either empty or contains all non-negative integers.
General cases highlights some properties for specific values of the two sets.
Finally, the case where both sets are equal and contains consecutive non-negative integers has been studied.
These sets are called segments.
It has revealed theoretical difficulties, especially when computing graph order between consecutive steps.
This has led to the creation of a new tool named the "sustainable interval".
This tool aims at estimating bounds that frames graph order even though it is not always reliable as probabilities are involved.
This last study is only about equal sets.
For further studies, the case where both sets are segments but not equal seems relevant as it does not change too much from the segment case.

\bibliographystyle{plain}       
\bibliography{biblio}   

\end{document}